\documentclass[11pt]{article}

\usepackage{amsmath}
\usepackage{amssymb}

\usepackage{graphicx}

\usepackage{cite}

\usepackage{color} 

\usepackage{setspace} 
\usepackage[labelfont=bf,labelsep=period,justification=raggedright]{caption}

\makeatletter
\renewcommand{\@biblabel}[1]{\quad#1.}
\makeatother

\date{}

\usepackage{lineno}
\usepackage{rotating}

\begin{document}
\title{Measuring Nepotism Through Shared Last Names: Response to Ferlazzo \& Sdoia}
\author{Stefano Allesina\\Department of Ecology \& Evolution and Computation Institute, University of Chicago,\\ Chicago IL 60637 USA. E-mail: sallesina@uchicago.edu}

\maketitle

\begin{abstract}
In a recent article, I showed that in several academic disciplines in
Italy, professors display a paucity of last names that cannot be
explained by unbiased, random, hiring processes. I suggested that this
scarcity of last names could be related to the prevalence of
nepotistic hires, i.e., professors engaging in illegal practices to
have their relatives hired as academics. My findings have recently
been questioned through repeat analysis to the United Kingdom
university system. Ferlazzo \& Sdoia found that several disciplines in
this system also display a scarcity of last names, and that a similar
scarcity is found when analyzing the first (given) names of Italian
professors. Here I show that the scarcity of first names in Italian
disciplines is completely explained by uneven male/female
representation, while the scarcity of last names in United Kingdom
academia is due to discipline-specific immigration. However, these
factors cannot explain the scarcity of last names in Italian
disciplines. Geographic and demographic considerations -- proposed as
a possible explanation of my findings -- appear to have no significant
effect: after correcting for these factors, the scarcity of last names
remains highly significant in several disciplines, and there is a
marked trend from north to south, with a higher likelihood of nepotism
in the south and in Sicily. Moreover, I show that in several Italian
disciplines positions tend to be inherited as with last names (i.e.,
from father to son, but not from mother to daughter). Taken together,
these results strenghten the case for nepotism, highlighting that
statistical tests cannot be applied to a dataset without carefully
considering the characteristics of the data and critically
interpreting the results.
\end{abstract}

\section*{Introduction}
Cases of nepotism have been documented in Italian universities
\cite{perotti2008universita}. Professors were found to engage in
illegal practices to have relatives hired as academics in their
institution or within the same discipline at other
universities. Because the prevalence of nepotism in Italian academia
is largely unknown, I recently proposed a statistical method to
estimate rates of nepotism based on the distribution of last names of
Italian academics \cite{allesina2011measuring}. My results confirmed
the findings obtained with more complex techniques
\cite{durante2011academic}, namely that a) some disciplines are more
likely to display nepotistic tendencies, and b) a latitudinal pattern
exists, showing a higher probability of nepotism in the south of the
country and in Sicily.

The method is simple and requires minimal data, only a list of the
names of all professors in Italy, and their corresponding
discipline. For a given discipline, count the number of academics $N$
and the number of unique last names $L$. Then draw a random sample
from the list of all Italian professors, taking $N$ professors without
repetition. Finally, compute the probability $p$ that a number of
unique last names $L' \leq L$ is found in a sample. If the probability
is very low, i.e., it is difficult to find a sample with a number of
last names smaller than that empirically observed for the discipline,
then the scarcity of last names in the discipline cannot be explained
by unbiased (i.e., random) hiring processes.

Clearly, the method does not measure nepotism per se
\cite{allesina2011measuring}, only whether the scarcity of last names
can or cannot be explained by a random process. The reasoning for
nepotism comes from the lack of alternative explanations for the
scarcity of last names. In my work I examined disciplines, which are
quite uniformly represented geographically, and thus geographical
considerations are unlikely to account for the scarcity of last name
in specific disciplines. Scarcity of last names could be due to
``career following'', the tendency of offspring to follow their
parents' careers, but I argued that this is quite unlikely
\cite{allesina2011measuring}, as even extremely specific fields of
research yield a significant scarcity of last names. For example,
``Philosophy and theory of languages'' has significantly fewer names
than expected at random \cite{allesina2011measuring}. Although the
sons and daughters of professors in this discipline could be attracted
towards philosophy or linguistics, the expectation that they will
naturally become professors of ``Philosophy and theory of languages''
is far fetched.

The study attracted much attention in Italy, receiving praise and
criticism in equal measure in the popular media. However, only the
study by Ferlazzo \& Sdoia \cite{Ferlazzo} addressed the results from
a scientific standpoint. In summary, Ferlazzo \& Sdoia criticized my
method and argued that neglecting to consider the regional
distribution of last names is likely to bias the conclusions. They
also repeated the analysis I proposed on a set of professors in the
United Kingdom, finding a scarcity of last names comparable to the
Italian case. Finally, they repeated the analysis of Italian
professors using first (given) names rather than last names, finding
that several disciplines ``somewhat surprisingly'' display a paucity
of first names. They concluded that the method I proposed cannot
measure nepotism, and that the results I presented are better
explained by ``social capital transfer'', or by geographic and
demographic considerations.

Here I show that social capital transfer, geography and demography do
not account for my results or those of Ferlazzo \& Sdoia. In fact, a)
the results obtained using Italian academics' last names are not
affected by the regional distribution of last names -- repeating the
analysis on a regional or macro-regional basis yields consistent
results, b) the scarcity of first names in some disciplines is
completely driven by skewed gender representation and immigration, and
c) immigration -- not social capital transfer -- is key to explaining
the United Kingdom result. Having ruled out the effects of immigration
and those stemming from the regional distribution of names, the
suspicion of nepotism, with a concentration in the south, is
strengthened, rather than weakened.

In addition, I show that academic positions in some disciplines tend
to be ``inherited'' as last names are. In Italy, women maintain their
maiden name, while sons and daughters only take the last name of their
father. As such, if we analyze only women, the potential for
discovering nepotism is greatly reduced (we can only find
inter-sibling nepotistic hires or those involving distant relatives);
on the other hand, when analyzing only men we have a higher potential
to discover nepotism, as pairs of fathers and sons included in the set
would share the same last name. Analyzing Italian academics divided by
gender I find no significant results with women, but highly
significant results with men: academic positions tend to be inherited
within families. Note that other geographic and demographic effects
would equally impact males and females.

The new analysis no only strengthens my findings, but highlights that
the proposed method -- as with any statistical procedure -- cannot be
applied to a dataset without first carefully considering the
characteristic of the data, and without critically interpreting the
results.

\section*{Results and Discussion}
\subsection*{Italy -- Last Names: Regional Analysis Confirms the Initial Findings}

In this section, I show that repeating the analysis I perfomed on the
whole set of Italian professors \cite{allesina2011measuring} using
only the professors working in a region or macro-region yields
comparable results: the scarcity of last names in some disciplines
cannot be explained by unbiased processes, and the likelihood of
nepotism is higher in the south and in Sicily.

Using the same data as in Allesina \cite{allesina2011measuring}, last
names have been transformed to upper case, and spaces and apostrophes
have been removed. I repeated the same analysis
\cite{allesina2011measuring} using: a) all academics; b) academics
divided into those working in the north, center, and south of Italy,
those working in Sardinia and those in Sicily; and c) academics
working in each of the 20 Italian regions. I present results obtained
using $10^5$ simulations for each case. I computed a p-value only for
disciplines with more than 50 professors (since for small disciplines
even significant differences would be negligible from a practical
standpoint). All code and data required for repeating the analysis
accompanies this article.

Throughout the article, I call highly significant results yielding a
p-value $\leq 0.05$ associated with a q-value $\leq 0.05$. That is,
the highly significant results have low probability of being obtained
at random (low p-value) and low probability of being false positives
(low q-value). I computed q-values using the {\tt R} package {\tt
  qvalue} using the bootstrap method \cite{qvalue}.

In Table 1 I report results for the regional analysis. When
considering all academics, 11 disciplines show a highly significant
scarcity of last names, with six having extremely small p-values $\leq
0.01$ (Industrial Engineering, Law, Medicine, Geography, Pedagogy and
Agriculture). I then divided the academics into macro-regions and
repeated the analysis. Clearly, using smaller samples necessarily
reduces the statistical power. Considering only the north (the regions
Aosta Valley, Liguria, Lombardy, Piedmont, Emilia-Romagna,
Friuli-Venezia Giulia, Trentino-Alto Adige and Veneto) and repeating
the analysis using the subset of professors working in this part of
the country I find a single discipline yields significant results
(Industrial Engineering). In the center (Lazio, Marche, Tuscany and
Umbria), Medicine and Chemistry yield a significant scarcity of last
names. Note that these three disciplines are included in the total
list of 11 that were significant at the national level. The results in
the south show a higher number of potentially nepotistic disciplines,
with 8 significant results (6 included in the list of 11 significant
at the national level). In Sicily 7 disciplines produce highly
significant results (5 in the list of the 11). In Sardinia, mainly due
to the small sample size, no result was highly significant. In
summary, disciplines found to be problematic at the national level are
among those problematic at the macro-regional level.

Repeating the analysis at the smaller, regional level further confirms
these findings. For each of the 20 Italian regions, I computed a
p-value for each discipline with more than 50 professors, and counted
how many results were $\leq 0.05$. Medicine yields low p-values in 8
regions out of 16, Industrial Engineering in 6 regions out of 15, Law
in 4 out of 17 (Table 1). If we order disciplines according to the
proportion of regions in which we find p-values $\leq 0.05$, we find
that the four with the highest proportions are included in the list of
the six with the lowest p-values at the national level.

The regional and macro-regional analysis of last names confirms the
findings of my previous work \cite{allesina2011measuring}. Some
disciplines have a highly significant scarcity of last names, which
can be interpreted as a higher likelihood of nepotism. The scarcity of
last names is more marked in the south of the country and in
Sicily. Note that the analysis has been conducted using a regional or
macro-regional pool of names, and thus is not compatible with the
explanation of Ferlazzo \& Sdoia who conjecture that the results could
be explained by ``migration [that] might have produced a larger
variability of last names in the northern regions and a lower
variability (i.e., more shared names) in the southern regions''
\cite{Ferlazzo}. These results are not surprising, as they closely
match the logistic regression analysis I performed
\cite{allesina2011measuring} and the study of Durante {\em et al.}
\cite{durante2011academic}: even when considering the local
distribution of last names, the scarcity of last names in some
disciplines cannot be explained.

\subsection*{Italy -- First Names: Gender Drives the Results}
In this section, I show that analyzing first names in Italian academia
simply highlights that in some disciplines women represent a small
minority of the professors.

\begin{figure}[!ht]
\begin{center}
\includegraphics[width=4in]{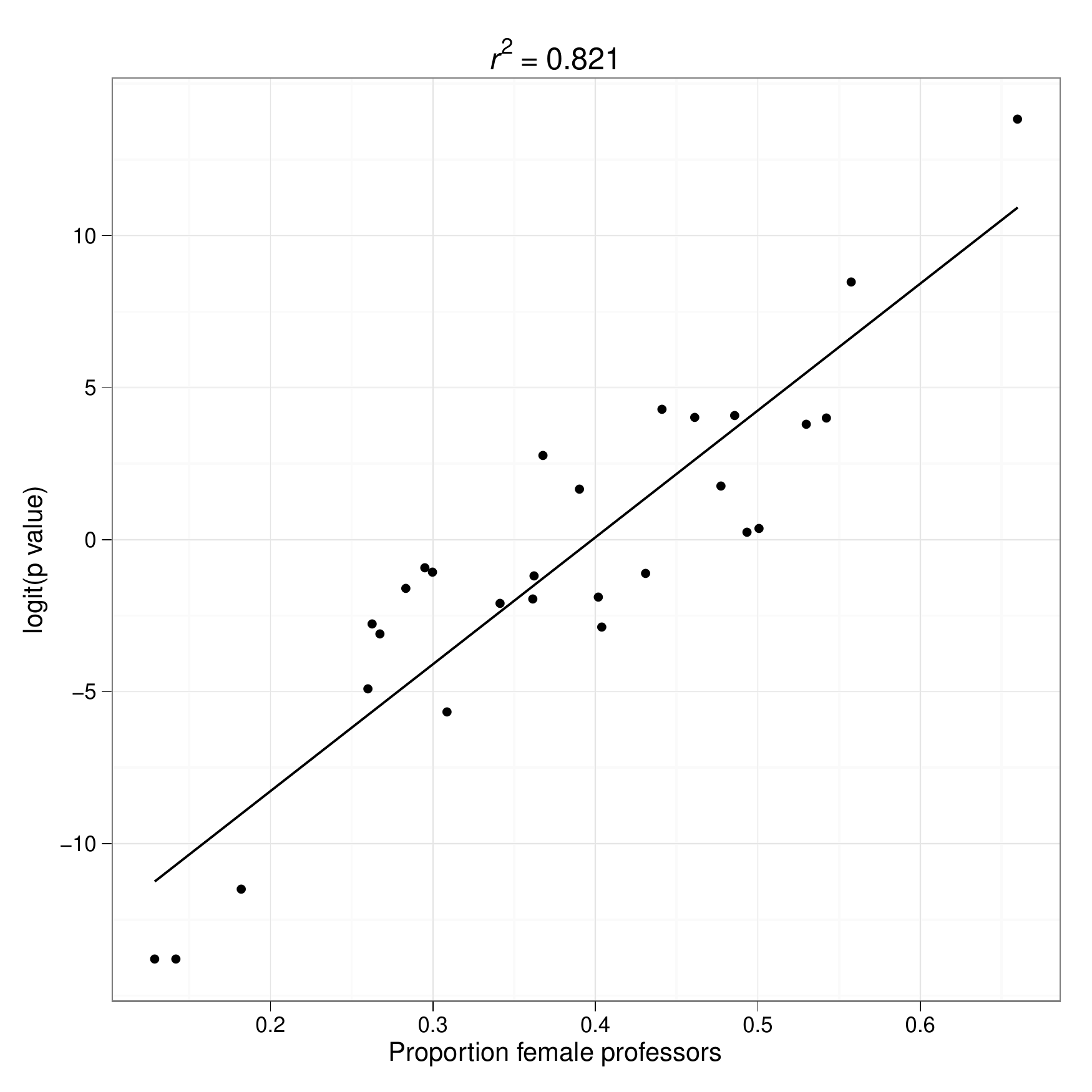}
\end{center}
\caption{ {\bf Gender Inequality and Scarcity of First Names.} Plot of
  $\text{logit}(p_i)=\log(p_i/(1-p_i))$ (y-axis), where $p_i$ is the
  p-value for discipline $i$ reported in Table 2 as a function of the
  fraction of women in the discipline. To avoid numerical problems,
  values $<10^{-6}$ have been set to $10^{-6}$ and those $>1 -
  10^{-6}$ to $1 - 10^{-6}$.  }
\label{FigureGenderFirst}
\end{figure}

The analysis of first names yields five disciplines with a highly
significant scarcity of first names (Table 2): Physics, Industrial and
Electronic Engineering, Economics and Earth Sciences have too few
first names compared to what would be expected at random. Linguistics
and Psychology, on the other hand, have a significant excess of first
names. Ferlazzo \& Sdoia find these results ``somewhat
surprising''. However, there is a well-documented bias in gender
representation in the so-called STEM (Science, Technology, Engineering
and Mathematics) disciplines, where professors are overwhelmingly
male. For example, in Italy more than 55\% of the professors in
Psychology are women, while in Electronic Engineering less that 13\%
are women. To test whether this observation can account for the
results, I divided the professors according to gender and computed the
proportion of women in each field. We can model the relationship
between gender representation and p-values as a logistic function: a
sigmoid describes the p-value as a function of the proportion of
women. If the fraction of women were driving the p-values, then
plotting $\text{logit}(\text{p-value})$ against the proportion of
women would produce a straight line. The plot indeed shows a strong
linear trend ($r^2>0.8$, Figure \ref{FigureGenderFirst}), indicating
that gender is likely to play a strong role in explaining the results.

To further test this hypothesis, I repeated the first-name analysis
dividing the professors according to gender. Among males, only
Electronic Engineering produces highly significant results. Among
females (representing about one third of the professors), four
disciplines are significant, meaning that gender can explain the
results for males but not for females.

In order to understand this new result, we have to recongnize a subtle
characteristic of the data. Immigration is unfortunately quite rare in
Italian academia, with the exception of a few disciplines. In
Linguistics, for example, native speakers are routinely hired to teach
languages. In Italy, immigration has little effect on the analysis of
last names, as there are very few immigrants and most last names are
rare even among the native Italians. Thus, the influx of rare foreign
last names has negligible effects. However, the analysis of first
names can be greatly impacted by immigration, as most Italian first
names are very common. Among all 61,340 professors, we find about
27,000 last names, but only about 7,000 first names. The most common
last name, ROSSI, is observed 225 times, while the most common first
name, GIUSEPPE, more than 1,400 times. Because immigrants represent a
source of unique or very rare first names, they could bias the results
by introducing unique first names in some disciplines but not in
others. The effect is likely to be larger for women, as there is one
distinct first name for every 5 male professors, but only one distinct
first name for every 7 women.

How can we remove immigrants from the analysis? Unfortunately,
information on the birthplace of Italian professors is not
available. To determine which disciplines are most affected by
immigration, I compiled a list of the 7,500 most common Italian last
names from phone book data (included with this article). I then
measured the proportion of common names in each discipline. Common
last names represent on average 52\% of the last names in each
discipline, with Linguistics (43\%) and Anthropology (42.6\%) being
the two disciplines with fewer common last names (bottom 5\% of the
distribution), and thus more likely to include many
immigrants. Removing the two disciplines (which are among the
smallest, representing about 3\% of the professors combined) yields
one significant result for each gender (M: Electronic Engineering, F:
Chemistry).

The scarcity of first names in some disciplines is therefore largely
explained by gender representation and immigration. Social capital
transfer, geographic or demographic effects appear to have no
appreciable effect. What is ``somewhat surprising'', therefore, is not
the result itself, but rather that Ferlazzo \& Sdoia failed to
consider gender as a potential explanation. In fact, countless studies
identified uneven gender representation in science and technology
\cite{hyde2008gender,hill2010so}, and several initiatives have been
implemented both in the European Union and in the United States to
solve this grave problem. A trivial prediction that arises from these
findings is that, when analyzing the distribution of first names in
any top institution, one would find -- unfortunately -- exactly the
same result: too few last names are present in the STEM disciplines,
because too few women work in these fields.

\subsection*{United Kingdom -- Last Names: Immigration Drives the Results}
In this section, I show that the results for the set of United Kingdom
professors can be explained by immigration, which affects some
disciplines more than others.

The dataset from the Research Assessment Exercise (RAE) 2008 is not as
well-suited for analysis as the Italian one. In fact, not all
professors are present, and each professor may be present more than
once -- following transfers between institutions
\cite{Ferlazzo}. Given that this latter characteristic could greatly
hamper the analysis of last names, I follow Ferlazzo \& Sdoia
\cite{Ferlazzo} by removing duplicate records identifiable as
professors with the same last name and initials (first names are not
available) working in the same discipline. Moreover, in the United
Kingdom it is very common for married women to take the last name of
the husband. This can be accomplished in a number of ways: the woman
can take the husband's last name, the two names could be hyphenated,
or the prefix ``nee''' could be added in front of the maiden
name. Because the dataset contains all types of composite names, I
kept only the first name in the case of hyphenated names (e.g.,
``PORCELLINI-SLAWINSKI''), and removed any name in parenthesis
(e.g.,``MCCARTHY (FORMERLY RIBBENS-MCCARTHY)'', ``NAUGHTON (NEE
LESNIEWSKA)''). This is important, as otherwise married women would
add unique names to the set, biasing the analysis, although Ferlazzo
\& Sdoia did not seem to consider this complication. Note that this is
not a problem in the Italian dataset, as women in Italy maintain their
maiden name.

In many countries ``spousal hires'' (i.e., hiring a couple at the same
time) are encouraged, and marriage -- with consequent change of last
name -- could happen after both partners have been hired. As such, in
these countries we expect the scarcity of last name to over-represent
nepotism, rather than greatly under-estimate it as in the Italian case
-- where spousal hires are explicitly forbidden and women keep their
maiden name.

I analyzed the 67 self-reported disciplines of the RAE (Units of
Assessment, Table 3). The results clearly signal that something other
than social capital transfer is responsible for the results. In fact,
24 out of 67 disciplines show a highly significant scarcity of last
names. Interestingly, Physics, Pure and Applied Mathematics, Computer
Science and Statistics all show an excess of last names, while Celtic
Studies, English Literature and History show a paucity of names. Why
would social capital transfer happen in some of the humanities, but
not in the mathematical sciences? If anything, we would expect the
opposite, as there have even been studies showing the heritability of
mathematical talent \cite{wijsman2004familial}, leading to the
expectation that career following should be more likely -- rather than
less -- in quantitative disciplines.

If we were to rule out career following, what else could explain the
pattern? To elucidate the results, we can start by listing the most
common name in each discipline of the Research Assesment Exercise. We
find that CHEN is the most common last name in Statistics, WANG in
Electric Engineering, in Mechanical Engineering, and in General
Engineering, while ZHANG dominates Asian Studies. These are all common
last names in China, while in the United Kingdom the most common names
are SMITH, JONES and TAYLOR. The large presence of professors of
Chinese descent in United Kingdom universities is due to the fact that
these institutions attract some of the best researchers from around
the world. However, not all the disciplines are equally impacted by
immigration, with foreigners more represented in the technical and
scientific fields. In comparison, all the most common names in Italian
disciplines sound undoubtedly Italian (with ROSSI, RUSSO and FERRARI
dominating, as in the general population).

To test whether immigration, rather than social capital transfer,
could explain the significant results, I took the 7,500 most common
last names in the United Kingdom (data included with the article) and
repeated the analysis using only the professors whose name is included
in the set. The rationale is that by analyzing only common names, the
relevance of social capital transfer would remain unchanged (we can
assume it is equally likely to happen regardless the last name), but
the effect of immigration, i.e., the influx of rare names, would be
greatly reduced. Clearly, the use of phone book data is not optimal,
as common immigrant last names are present (e.g., GALLO, RICCI and
ROSSI, common Italian last names, the common Indian last names NAIR,
SHARMA, SINGH, PATEL, and the common Chinese CHANG, CHONG, LING, TANG,
WONG, YANG are all present in the 7,500 names). Moreover, we are
removing the last names of native professors when rare, reducing
statistical power. However, in the absence of data on the immigration
status of the academics, this analysis serves as a reasonable first
approximation.

Repeating the analysis using only the 7,500 ``common'' names yield
only three highly significant results instead of the 24 obtained when
using all names (two included in the original list of 24). Thus,
immigration largely accounts for the observed pattern.

\subsection*{Analysis of Italian Last Names -- Effects of Immigration and Gender}
In this section, I show that the results obtained for the Italian last
names are robust to the removal of rare names, and that academic
positions tend to be inherited within families.

To show that the results in the Italian case are robust, I repeated
the analysis of last names using only the 7,500 most common Italian
last names (Table 4). Of the six disciplines yielding the lowest
p-values when analyzing all names, all but Agriculture are still
highly significant. Hence, although there has been a reduction in the
number of highly significant disciplines, it is much less marked than
in the United Kingdom case.

Analyzing the last names by gender provides an interesting test. In
fact, because in Italy only the last name of the father is passed on
to sons and daughters, the potential for nepotism is greatly reduced
when analyzing only women. Among women, only intra-sibling nepotistic
hires or those involving distant relatives would be recorded, while
among men father-son relationships would also be counted. Besides this
difference, the geographical and demographic distribution of last
names should affect men and women equally, providing a strong test for
the absence of these effects. I therefore repeated the analysis of
last names dividing professors into men and women. The results provide
strong support for the ``inheritability'' of academic positions within
families (Table 4). In fact, four disciplines (Medicine, Law,
Industrial Engineering and Political Science) yield a significant
scarcity of last names when considering only men, while no discipline
is significant when considering only women.

\section*{Conclusions}
In my previous article \cite{allesina2011measuring}, I proposed a
method to determine whether the scarcity of names in a discipline
could be determined by unbiased processes. This is not the case for
some disciplines in Italy when analyzing first or last names, and in
the United Kingdom analyzing last names. However, the cause for the
scarcity of names differs between these cases.

When analyzing Italian first names, I showed that the fraction of
women in each discipline strongly correlates with the significance
(p-value). Hence, my method -- as applied by Ferlazzo \& Sdoia
\cite{Ferlazzo} -- can be seen as a computationally-intensive
technique to measure the obvious gender representation bias in,
typically, STEM disciplines. Note that in these disciplines there
really are fewer first names than one would expect, but the problem
this result is highlighting -- the scarcity of women in STEM
disciplines -- is much more difficult to solve than nepotism, and
unfortunately affects academic institutions worldwide.

The analysis of last names is better suited to systems where
immigration is negligible, as the Italian case. When applied to a
system where immigration is common, and more common in some
disciplines than others, the scarcity of last names signals the
differential impact of immigration, rather than nepotism. 

These findings stress that statistical tests cannot be applied to a
dataset without considering the special characteristics of the data
and critically interpreting the results. Many of the conjectures in
Ferlazzo \& Sdoia \cite{Ferlazzo} could have been tested quite easily,
as shown here. And in fact, testing for the absence of geographical
and demographic considerations make the case for nepotism stronger.

In Italian academia, some disciplines, especially Medicine, Law and
Industrial Engineering, yield a scarcity of last names that cannot be
explained neither by geographic nor demographic considerations, a
finding that is also robust to analysis of common names only. When
testing only men or women, I showed that positions tend to be
``inherited'' as with last names, i.e., from father to son. All these
considerations lead to the conclusion that nepotism is the most
logical explanation for these findings.


\section*{Acknowledgments}
P. Staniczenko, A. Ekl\"of, P. Lemos \& S. Tang for comments and
discussion.

\bibliography{ResponseNepo}

\section*{Figures and Tables}

\begin{sidewaystable}[!ht]
  \caption{
    \bf{Analysis of Italian Academia: Last Names}}
  \small
  \begin{tabular}{|r|l|c|c|c|c|c|c|c|c|c|c|}
    \hline
    Discipline & Code & Academics & Last Names & p & North-p & Center-p & South-p & Sardinia-p & Sicily-p & Prop. Regions & Count\\
    \hline
    Industrial Eng. & ING-IND & 3180 & 2686 & \underline{$<$0.001} & \underline{0.002} & 0.135 & \underline{$<$0.001} & 0.175 & 0.038 & 0.4 & 6/15\\
    Law & IUS & 5144 & 4023 & \underline{$<$0.001} & 0.523 & 0.043 & \underline{$<$0.001} & 0.035 & \underline{$<$0.001} & 0.235 & 4/17\\
    Medical Sciences & MED & 10783 & 7461 & \underline{$<$0.001} & 0.014 & \underline{$<$0.001} & \underline{$<$0.001} & 0.013 & 0.075 & 0.5 & 8/16\\
    Geography & M-GGR & 377 & 359 & \underline{0.005} & 0.718 & 0.444 & \underline{0.008} & - & - & - & 0/0\\
    Pedagogy & M-PED & 675 & 634 & \underline{0.005} & 0.285 & 0.08 & 0.06 & - & \underline{$<$0.001} & 0.167 & 1/6\\
    Agriculture & AGR & 2345 & 2057 & \underline{0.008} & 0.233 & 0.012 & 0.081 & 0.197 & \underline{0.003} & 0.286 & 4/14\\
    Civil Eng. & ICAR & 3836 & 3204 & \underline{0.01} & 0.606 & 0.066 & \underline{0.007} & 0.388 & 0.099 & 0.25 & 4/16\\
    Mathematics & MAT & 2531 & 2211 & \underline{0.019} & 0.116 & 0.937 & 0.057 & - & \underline{$<$0.001} & 0.154 & 2/13\\
    Chemistry & CHIM & 3129 & 2683 & \underline{0.036} & 0.055 & \underline{0.005} & 0.159 & 0.298 & \underline{$<$0.001} & 0.188 & 3/16\\
    Earth Sciences & GEO & 1196 & 1105 & \underline{0.047} & 0.134 & 0.482 & 0.402 & - & 0.068 & 0 & 0/9\\
    Philosophy & M-FIL & 1125 & 1043 & \underline{0.048} & 0.29 & 0.792 & \underline{0.021} & - & 0.067 & 0.222 & 2/9\\
    History & M-STO & 1453 & 1329 & 0.061 & 0.537 & 0.268 & 0.677 & - & 0.698 & 0.1 & 1/10\\
    Statistics & SECS-S & 1212 & 1122 & 0.088 & 0.647 & 0.145 & \underline{0.002} & - & 0.032 & 0.5 & 4/8\\
    Veterinary & VET & 847 & 799 & 0.126 & 0.604 & 0.127 & 0.07 & - & \underline{$<$0.001} & 0.222 & 2/9\\
    Political Sciences & SPS & 1792 & 1622 & 0.14 & 0.02 & 0.241 & 0.373 & 0.756 & 0.98 & 0.077 & 1/13\\
    Life Sciences & BIO & 5140 & 4172 & 0.153 & 0.058 & 0.569 & 0.368 & 0.135 & 0.527 & 0.125 & 2/16\\
    Informatics & INF & 834 & 789 & 0.192 & 0.136 & 0.07 & 0.472 & - & - & 0.143 & 1/7\\
    Physics & FIS & 2472 & 2186 & 0.245 & 0.475 & 0.272 & 0.476 & - & 0.34 & 0 & 0/13\\
    Philology & L-FIL-LET & 1780 & 1618 & 0.278 & 0.709 & 0.464 & 0.104 & - & 0.813 & 0.083 & 1/12\\
    Economics & SECS-P & 3806 & 3221 & 0.291 & 0.363 & 0.115 & \underline{0.005} & 0.033 & \underline{0.003} & 0.235 & 4/17\\
    Physical Ed. & M-EDF & 138 & 136 & 0.347 & - & - & - & - & - & - & 0/0\\
    Art History & L-ART & 815 & 775 & 0.36 & 0.175 & 0.771 & 0.289 & - & - & 0 & 0/7\\
    Electronic Eng. & ING-INF & 2089 & 1880 & 0.389 & 0.266 & 0.604 & 0.128 & - & 0.439 & 0 & 0/12\\
    Archeology & L-ANT & 704 & 678 & 0.702 & 0.319 & 0.593 & 0.302 & - & 0.664 & 0 & 0/6\\
    Near Eastern & L-OR & 317 & 312 & 0.741 & 0.725 & 1 & 1 & - & - & 0 & 0/2\\
    Psichology & M-PSI & 1252 & 1176 & 0.769 & 0.325 & 0.611 & 0.794 & - & 0.23 & 0 & 0/8\\
    Linguistics & L-LIN & 2173 & 2000 & 1 & 0.983 & 0.998 & 0.909 & 0.61 & 0.215 & 0 & 0/15\\
    Demography & M-DEA & 195 & 195 & 1 & 1 & 1 & - & - & - & - & 0/0\\
    \hline
  \end{tabular}
  \normalsize
  \begin{flushleft}
    For each discipline, I report the number of academics, the number
    of last names and the p-value obtained considering all
    professors. For each macro-region (North, Center, South, Sardinia,
    Sicily), I report the p-value obtained repeating the analysis on
    the subset of professor working in the macro region. I also report
    the proportion of regions in which a discipline yields a p-value
    $\leq 0.05$ and the count of regions with low p-values. In fact, I
    only considered regions in which 50 or more academics are working
    in a given discipline. Underlined (highly significant) values are
    those with p-value and q-value $\leq 0.05$.
  \end{flushleft}
  \label{tab:italast}
\end{sidewaystable}

\begin{table}[!ht]
  \caption{
    \bf{Analysis of Italian Academia: First Names}}
  \small
  \begin{tabular}{|l|c|c|c|c|c|c|c|}
    \hline
    Code & Academics & First Names & p & F-p & M-p & F-p' & M-p' \\
    \hline
    ING-INF & 2089 & 519 & \underline{$<$0.001} & \underline{0.012} & \underline{$<$0.001} & 0.019 & \underline{$<$0.001}\\
    ING-IND & 3180 & 762 & \underline{$<$0.001} & 0.058 & 0.015 & 0.095 & 0.04\\
    FIS & 2472 & 685 & \underline{$<$0.001} & 0.196 & 0.204 & 0.279 & 0.316\\
    SECS-P & 3806 & 950 & \underline{0.003} & 0.031 & 0.104 & 0.096 & 0.2\\
    GEO & 1196 & 439 & \underline{0.007} & 0.427 & 0.083 & 0.504 & 0.119\\
    ICAR & 3836 & 975 & 0.042 & 0.546 & 0.239 & 0.739 & 0.395\\
    CHIM & 3129 & 853 & 0.053 & \underline{0.001} & 0.403 & \underline{0.005} & 0.542\\
    INF & 834 & 354 & 0.058 & 0.56 & 0.264 & 0.618 & 0.325\\
    IUS & 5144 & 1200 & 0.108 & 0.353 & 0.112 & 0.665 & 0.233\\
    MAT & 2531 & 749 & 0.124 & \underline{0.013} & 0.537 & 0.038 & 0.662\\
    SECS-S & 1212 & 461 & 0.13 & 0.08 & 0.148 & 0.131 & 0.196\\
    MED & 10783 & 2002 & 0.165 & 0.334 & 0.683 & 0.771 & 0.913\\
    M-EDF & 138 & 99 & 0.231 & - & 0.089 & - & 0.095\\
    VET & 847 & 368 & 0.244 & \underline{0.024} & 0.409 & 0.041 & 0.468\\
    AGR & 2345 & 721 & 0.253 & 0.981 & 0.056 & 0.993 & 0.1\\
    M-FIL & 1125 & 446 & 0.281 & 0.375 & 0.646 & 0.456 & 0.72\\
    M-GGR & 377 & 218 & 0.559 & 0.209 & 0.305 & 0.247 & 0.331\\
    BIO & 5140 & 1234 & 0.586 & \underline{0.004} & 0.794 & 0.051 & 0.894\\
    M-STO & 1453 & 551 & 0.839 & 0.414 & 0.809 & 0.547 & 0.864\\
    L-ART & 815 & 379 & 0.851 & 0.526 & 0.425 & 0.624 & 0.48\\
    SPS & 1792 & 640 & 0.941 & 0.333 & 0.982 & 0.474 & 0.991\\
    L-ANT & 704 & 355 & 0.978 & 0.206 & 0.967 & 0.276 & 0.975\\
    M-PED & 675 & 346 & 0.982 & 0.685 & 0.677 & 0.759 & 0.712\\
    L-FIL-LET & 1780 & 646 & 0.982 & 0.658 & 0.834 & 0.804 & 0.887\\
    L-OR & 317 & 208 & 0.983 & 0.788 & 0.8 & 0.816 & 0.821\\
    M-DEA & 195 & 147 & 0.986 & 0.789 & 0.929 & - & -\\
    M-PSI & 1252 & 536 & 1 & 0.55 & 0.997 & 0.701 & 0.998\\
    L-LIN & 2173 & 944 & 1 & 1 & 1 & - & -\\
    \hline
  \end{tabular}
  \normalsize
  \begin{flushleft}
    For each discipline, I report the number of academics, the number
    of first names and the p-value obtained considering the first
    names of all professors. I then computed the p-values for the
    first names within each discipline when I divide the professors in
    female (F-p) and male (M-p). Finally, I compute the p-values when
    the professors in Linguistics (L-LIN) and Demography and Ethnology
    (M-DEA) have been removed (F-p', M-p'). These two fields contain
    the highest proportion of foreigners, providing a source of unique
    or very rare first names.
  \end{flushleft}
  \label{tab:itafirst}
\end{table}

\begin{table}[!ht]
  \caption{
    \bf{Analysis of UK Academia: Last Names}}
  \tiny
  \begin{tabular}{|r|l|c|c|c|c|}
    \hline
    Discipline & Code & Academics & Last Names & p & Common-p\\
    \hline
    Chemistry & 18 & 1252 & 987 & \underline{$<$0.001} & \underline{$<$0.001}\\
    Celtic Studies & 56 & 158 & 138 & \underline{$<$0.001} & \underline{$<$0.001}\\
    English Language and Literature & 57 & 2343 & 1682 & \underline{$<$0.001} & 0.034\\
    Sports-Related Studies & 46 & 585 & 501 & \underline{$<$0.001} & 0.037\\
    History & 62 & 2296 & 1721 & \underline{$<$0.001} & 0.038\\
    Education & 45 & 2232 & 1639 & \underline{$<$0.001} & 0.054\\
    Geography and Environmental Studies & 32 & 1442 & 1133 & \underline{$<$0.001} & 0.074\\
    Nursing and Midwifery & 11 & 816 & 678 & \underline{$<$0.001} & 0.137\\
    Agriculture, Veterinary and Food Science & 16 & 1227 & 991 & \underline{$<$0.001} & 0.175\\
    Social Work and Social Policy & 40 & 1616 & 1260 & \underline{$<$0.001} & 0.464\\
    Archaeology & 33 & 668 & 577 & \underline{0.001} & 0.041\\
    Earth and Environmental Sciences & 17 & 1423 & 1153 & \underline{0.001} & 0.184\\
    Dentistry & 10 & 471 & 417 & \underline{0.003} & 0.04\\
    Biological Sciences & 14 & 2785 & 2092 & \underline{0.003} & 0.144\\
    Health Services Research & 7 & 637 & 554 & \underline{0.003} & 0.218\\
    Epidemiology and Public Health & 6 & 711 & 613 & \underline{0.003} & 0.243\\
    Pharmacy & 13 & 550 & 484 & \underline{0.004} & 0.062\\
    Mechanical and Aeronautical Engineering & 28 & 1149 & 956 & \underline{0.005} & 0.326\\
    Town and Country Planning & 31 & 530 & 469 & \underline{0.008} & 0.116\\
    Music & 67 & 777 & 670 & \underline{0.009} & 0.154\\
    Asian Studies & 49 & 185 & 172 & \underline{0.02} & 0.062\\
    Drama, Dance and Performing Arts & 65 & 574 & 508 & \underline{0.02} & 0.12\\
    Other Hospital Based Clinical Subjects & 4 & 1935 & 1535 & \underline{0.026} & 0.48\\
    Iberian and Latin American Languages & 55 & 278 & 256 & \underline{0.029} & 0.023\\
    General Engineering and Mining Engineering & 25 & 1817 & 1458 & 0.055 & 0.175\\
    Allied Health Professions and Studies & 12 & 1803 & 1449 & 0.063 & 0.825\\
    Theology, Divinity and Religious Studies & 61 & 638 & 565 & 0.063 & 0.287\\
    Classics & 59 & 558 & 499 & 0.069 & 0.073\\
    Civil Engineering & 27 & 635 & 563 & 0.07 & 0.367\\
    Chemical Engineering & 26 & 258 & 240 & 0.073 & 0.133\\
    Sociology & 41 & 1232 & 1033 & 0.076 & 0.726\\
    Electrical and Electronic Engineering & 24 & 997 & 854 & 0.088 & 0.118\\
    Psychiatry and Neuroscience & 9 & 956 & 822 & 0.093 & 0.043\\
    Communication, Cultural and Media Studies & 66 & 688 & 608 & 0.099 & 0.567\\
    Art and Design & 63 & 2356 & 1837 & 0.101 & 0.933\\
    Metallurgy and Materials & 29 & 500 & 452 & 0.112 & 0.26\\
    Primary Care & 8 & 197 & 186 & 0.131 & 0.672\\
    French & 52 & 485 & 440 & 0.135 & 0.665\\
    Philosophy & 60 & 742 & 655 & 0.17 & 0.513\\
    German, Dutch and Scandinavian Languages & 53 & 280 & 262 & 0.184 & 0.048\\
    Linguistics & 58 & 407 & 375 & 0.23 & \underline{0.004}\\
    Psychology & 44 & 1977 & 1587 & 0.256 & 0.838\\
    Library and Information Management & 37 & 363 & 337 & 0.261 & 0.672\\
    Anthropology & 42 & 464 & 426 & 0.321 & 0.192\\
    Cancer Studies & 2 & 901 & 788 & 0.329 & 0.446\\
    Law & 38 & 1987 & 1601 & 0.396 & 0.319\\
    History of Art, Architecture and Design & 64 & 459 & 423 & 0.401 & 0.579\\
    Infection and Immunology & 3 & 784 & 697 & 0.463 & 0.462\\
    Accounting and Finance & 35 & 212 & 203 & 0.464 & 0.588\\
    Middle Eastern and African Studies & 48 & 201 & 193 & 0.493 & 0.476\\
    Russian, Slavonic and East European Languages & 51 & 172 & 166 & 0.501 & 0.173\\
    Development Studies & 43 & 277 & 263 & 0.512 & 0.383\\
    American Studies and Anglophone Area Studies & 47 & 103 & 101 & 0.624 & 0.706\\
    Politics and International Studies & 39 & 1635 & 1358 & 0.637 & 0.461\\
    Pre-clinical and Human Biological Sciences & 15 & 722 & 650 & 0.641 & 0.484\\
    European Studies & 50 & 576 & 528 & 0.674 & 0.401\\
    Architecture and the Built Environment & 30 & 740 & 667 & 0.726 & 0.596\\
    Italian & 54 & 140 & 137 & 0.732 & 1\\
    Business and Management Studies & 36 & 3999 & 2945 & 0.761 & 0.007\\
    Statistics and Operational Research & 22 & 430 & 404 & 0.81 & 0.857\\
    Cardiovascular Medicine & 1 & 466 & 437 & 0.866 & 0.907\\
    Physics & 19 & 2072 & 1690 & 0.914 & 0.295\\
    Other Laboratory Based Clinical Subjects & 5 & 328 & 317 & 0.977 & 0.989\\
    Pure Mathematics & 20 & 793 & 723 & 0.978 & 0.106\\
    Applied Mathematics & 21 & 984 & 885 & 0.995 & 0.08\\
    Computer Science and Informatics & 23 & 2144 & 1794 & 1 & 0.304\\
    Economics and Econometrics & 34 & 1075 & 982 & 1 & 0.259\\
    \hline
  \end{tabular}
  \normalsize
  \newpage
  \begin{flushleft}
    For each discipline, I report the number of academics, the number
    of last names and the p-value obtained considering the last names
    of all professors. I then computed the p-values when only the
    7,500 most common last names are considered.
  \end{flushleft}
  \label{tab:uklast}
\end{table}

\begin{table}[!ht]
  \caption{
    \bf{Analysis of Italian Academia: Common Last Names and Gender}}
  \small
  \begin{tabular}{|r|c|c|c|c|}
    \hline
    Code & p & Common-p & F-p & M-p\\
    \hline
    MED & \underline{$<$0.001} & \underline{$<$0.001} & 0.044 & \underline{$<$0.001}\\
    ING-IND & \underline{$<$0.001} & \underline{0.003} & 0.453 & \underline{0.001}\\
    IUS & \underline{$<$0.001} & \underline{0.006} & 0.022 & \underline{$<$0.001}\\
    M-GGR & \underline{0.005} & \underline{0.028} & 0.38 & 0.213\\
    M-PED & \underline{0.005} & \underline{0.036} & 0.214 & 0.044\\
    AGR & \underline{0.008} & 0.27 & 0.126 & 0.361\\
    ICAR & \underline{0.01} & 0.104 & 0.082 & 0.02\\
    MAT & \underline{0.019} & 0.289 & 0.005 & 0.272\\
    CHIM & \underline{0.036} & 0.149 & 0.146 & 0.153\\
    GEO & \underline{0.047} & 0.178 & 0.26 & 0.193\\
    M-FIL & \underline{0.048} & 0.126 & 0.375 & 0.259\\
    M-STO & 0.061 & 0.271 & 0.61 & 0.1\\
    SECS-S & 0.088 & 0.258 & 0.832 & \underline{$<$0.001}\\
    VET & 0.126 & 0.403 & 0.208 & 0.249\\
    SPS & 0.14 & 0.427 & 0.036 & 0.741\\
    BIO & 0.153 & 0.751 & 0.55 & 0.065\\
    INF & 0.192 & 0.539 & 0.364 & 0.382\\
    FIS & 0.245 & 0.469 & 0.25 & 0.56\\
    L-FIL-LET & 0.278 & 0.447 & 0.045 & 0.848\\
    SECS-P & 0.291 & 0.716 & 0.892 & 0.076\\
    M-EDF & 0.347 & 0.342 & - & 0.399\\
    L-ART & 0.36 & 0.515 & 0.701 & 0.063\\
    ING-INF & 0.389 & 0.529 & 0.306 & 0.502\\
    L-ANT & 0.702 & 0.664 & 0.85 & 0.965\\
    L-OR & 0.741 & 0.667 & 1 & 0.816\\
    M-PSI & 0.769 & 0.61 & 0.794 & 0.463\\
    L-LIN & 1 & 0.335 & 0.98 & 0.978\\
    M-DEA & 1 & 1 & 1 & 1\\
    \hline
 \end{tabular}
  \normalsize
  \newpage
  \begin{flushleft}
    For each discipline, I report the p-value obtained considering the
    last names of all professors. I then computed the p-values when
    only the 7,500 most common last names are considered, and when
    professors are divided according to gender.
  \end{flushleft}
  \label{tab:itacommon}
\end{table}

\end{document}